\documentclass[conference]{IEEEtran}
\IEEEoverridecommandlockouts
\usepackage{cite}
\usepackage{multicol}
\usepackage{amsmath,amssymb,amsfonts}
\usepackage{algorithmic}
\usepackage{graphicx}
\usepackage{textcomp}
\usepackage{xcolor}
\def\BibTeX{{\rm B\kern-.05em{\sc i\kern-.025em b}\kern-.08em
    T\kern-.1667em\lower.7ex\hbox{E}\kern-.125emX}}

\def\lit{LIT}
\def\dlite{DLITE} 
\def\dl{DL} 

\begin{document}

\title{On Triangular Inequality of the Discounted Least Information Theory of Entropy (DLITE)\\
}

\author{
    \IEEEauthorblockN{Kashti S. Umare}
    \IEEEauthorblockA{
        \textit{Mathematics} \\
        \textit{Downingtown STEM Academy}\\
        Downingtown, U.S.A \\
        kashtiumare@gmail.com
    }
    \and
    \IEEEauthorblockN{Weimao Ke}
    \IEEEauthorblockA{
        \textit{College of Computing \& Informatics} \\
        \textit{Drexel University}\\
        Philadelphia, U.S.A. \\
        wk@drexel.edu
    }
}

\maketitle

\begin{abstract}
The Discounted Least Information Theory of Entropy (DLITE) is a new information measure that quantifies the amount of entropic difference between two probability distributions \cite{Ke2020}. It manifests multiple critical properties both as an information-theoretic quantity and as metric distance. In the report, we provide a proof of the triangular inequality of DLITE's cube root ($\sqrt[3]{DL}$), an important property of a metric, along with alternative proofs for two additional properties.
\end{abstract}


\section{Introduction}

Information and probability theories provide guidance in many fields such as information retrieval (IR), e.g. for the development of probabilistic retrieval and language modeling \cite{Robertson2009}. Shannon's information-theoretic entropy lays the foundation to quantify the amount of information in probability distributions and has a wide range of applications in computing and data sciences \cite{Shannon1948,Shaw:entropy}.

Based on Shannon entropy, information measures such as Kullback-Leibler (KL) divergence (relative entropy) and mutual information enable the computation of certain ``distance'' between probability distributions and give rise to classic term weighting schemes such as the Inverse Document Frequency (IDF) \cite{Kullback:1959,Amati:2002,Aizawa:2003}. IDF quantifies the amount of information due to a term (word) in a specific document by measuring its KL divergence from the collection-wide probability distribution.

However, KL divergence is not a metric and cannot be used as a symmetric distance measure. In addition, KL is unbounded and has undesirable consequences in practical applications, where extremely large value can dominate the scoring function\cite{Ke2013JCDL,Ke2015IJDL2}. Research has proposed the Discounted Least Information Theory (DLITE) as an alternative to mitigate some of these issues\cite{Ke2020}. It is observed that DLITE is bounded and possesses metric properties, including triangular inequality. In this report, we offer a proof of the triangular inequality of DLITE's cube root ($\sqrt[3]{DL}$) along with alternative proofs for two additional properties.

\section{{\dlite} Theory}

The Discounted Least Information Theory of Entropy (DLITE) is an extension of a prior work on the Least Information Theory (LIT) that satisfies several additional properties as an information metric. We shall introduce the LIT measure first.

\subsection{{\lit} Measure}

The Least Information Theory ({\lit}) quantifies the amount of entropic difference between two probability distributions \cite{Ke2013JCDL}. Given distributions $P$ and $Q$ of variable $X$, {\lit} is computed by:

\begin{eqnarray}
  {\lit}(P, Q)
  & = & \sum_{x \in X} \int_{p_x}^{q_x} - \log p\ dp \\
  & = & \sum_{x \in X} \Big\lvert p_x (1-\ln p_x) - q_x (1-\ln q_x) \Big\rvert
\end{eqnarray}

where $x$ is one of the mutually exclusive inferences of $X$, and $p_x$ and $q_x$ are probabilities of $x$ on the $P$ and $Q$ distributions respectively.

For any probabilities $p$ and $q$, let:

\begin{eqnarray}
g(p, q) & = & \Big\lvert p (1-\ln p) - q (1-\ln q) \Big\rvert
\end{eqnarray}

{\lit} can be written as:

\begin{eqnarray}
  {\lit}(P, Q)
  & = & \sum_{x \in X} g(p_x, q_x)
\end{eqnarray}

Research has applied LIT to data clustering, classification, and information retrieval, and shown its competitive performances compared to classic baselines\cite{Ke2015IJDL2,Ke2017,Du2018}.

\subsection{Entropy Discount}

For DLITE, the following entropy discount is introduced:

\begin{eqnarray}
  \Delta_H(P, Q)
  & = & \sum_{x \in X} \Big\lvert p_x - q_x \Big\rvert \frac{\int_{p_x}^{q_x} - p \log{p}\ dp}{\int_{p_x}^{q_x} x\ d x} \\
  & = & \sum_{x \in X} \frac{\Big\lvert p_x^2 (1 - 2\ln{p_x}) - q_x^2 (1-2\ln{q_x}) \Big\rvert}{2(p_x + q_x)}
\end{eqnarray}

For any probabilities $p$ and $q$, let:

\begin{eqnarray}
\delta_h(p, q) & = & \frac{\Big\lvert p^2 (1 - 2\ln{p}) - q^2 (1-2\ln{q}) \Big\rvert}{2(p + q)}
\end{eqnarray}

The entropy discount $\Delta_H$ can be written as:

\begin{eqnarray}
  \Delta_H(P, Q) & = & \sum_{x \in X} \delta_h(p_x, q_x)
\end{eqnarray}

\subsection{{\dlite}: {\lit} with Entropy Discount}

We now define the Discounted Least Information Theory of Entropy, or {\dlite}, as the amount of {\lit} subtracted by its entropy discount $\Delta_H$:

\begin{eqnarray}
{\dl}(P, Q)
& = & {\lit}(P, Q) - \Delta_H(P, Q) \\
& = & \sum_{x \in X} g(p_x, q_x) - \delta_h(p_x, q_x)
\label{eq:dl}
\end{eqnarray}

For any probability change from $p$ to $q$, let:

\begin{eqnarray}
 dl(p, q) & = & g(p, q) - \delta_h(p, q)
\label{eq:dl_x}
\end{eqnarray}

Equation~\ref{eq:dl} can written as:

\begin{eqnarray}
{\dl}(P, Q) & = & \sum_{x \in X}  dl(p_x, q_x)
\label{eq:dl_sum}
\end{eqnarray}

\subsection{{\dlite} Properties}

Again, {\dlite} is the amount of {\lit} with the $\Delta_H$ discount:

\begin{eqnarray}
  {\dl}(P, Q)
  & = & {\lit}(P, Q) - \Delta_H(P, Q) \\
  & = &  \sum_{x \in X} \int_{p_x}^{q_x} \log{\frac{1}{p}}\ dp \\
  &   & - \sum_{x \in X} \Big\lvert p_x - q_x \Big\rvert  \frac{\int_{p_x}^{q_x} p \log{\frac{1}{p}}\ dp}{\int_{p_x}^{q_x} p\ d p}
\label{eq:dl2}
\end{eqnarray}

Whereas {\lit} represents the sum of weighted, microscopic entropy changes, it consists of an amount of entropy change due to the scale of related probabilities, leading to an undesirable consequence of having different {\lit} amounts in different sub-system breakdowns. The entropy discount, $\Delta_H$, accounts for this extra amount in the {\lit} and reduces it to a scale-free measure. As shown in Equation~\ref{eq:dl2}, the discount on each $x$ dimension is a product of the absolute probability change in $p$ and a mean of $\log{\frac{1}{p}}$.

We discussed justifications of DLITE with a list of metric and information-theoretical properties in \cite{Ke2020}. We highlight DLITE's major theoretical properties below.

\subsubsection*{Metric Properties}

Given the definition in Equation~\ref{eq:dl} or \ref{eq:dl2}, it can be shown that {\dlite} satisfies the following metric properties:

\begin{enumerate}
  \item Non-negativity: ${\dl}(P,Q) \ge 0$ for any probability distributions $P$ and $Q$ of the same dimensionality. See Appendix for proof of Theorem 1.
  \item Identity of Indiscernibles: ${\dl}(P,Q) = 0$ if and only if $P$ and $Q$ are identical distributions.
  \item Symmetry: ${\dl}(P,Q) == {\dl}(Q, P)$, the amount of the information from $P$ to $Q$ is the same as that from $Q$ to $P$.
  \item Triangular Inequality: The cube root of DLITE satisfies the triangular inequality, that is $\sqrt[3]{ dl(p,q)} + \sqrt[3]{ dl(q,r)} \ge \sqrt[3]{ dl(p,r)}$.
\end{enumerate}

The focus of this report is to provide the proof for the 4th property above (Theorem 2 in the next section), along with alternative proofs for the 1st metric property (Theorem 1) and the following information-theoretic property (a lemma):

\begin{itemize}
    \item {\dlite} of an ensemble is the weighted sum of {\dlite}s in its sub-systems, $dl(xp, xq)=x\cdot dl(p,q)$.
\end{itemize}

\section{Theorems and Proofs}

\subsection{$1^{st}$ Metric Property: Non-negativity}

\setlength\parindent{24pt}
\textbf{Theorem 1.} \emph{For any probability distributions P and Q:}

\centering ${DL(P,Q)} \ge 0$

\raggedright
\emph{Proof.}
First we assume for all $p \in P$ and $q \in Q$ we have x is the larger of $p$ and $q$ and c is the smaller without loss of generality. Thus $x \geq c$. From here we know: \\

 \bigskip
\begin{center}
$dl(x, c) = lit(x,c) - \delta(x,c)$ \\

\bigskip

$= x(1-ln(x)) - c(1-ln(c)) - \frac{x^2(1-2ln(x))-c^2(1-2ln (c))}{2(x+c)}$

\end{center}

Note here that when $x=c$ is the only point where our function is equal to 0. With this understanding we can take the derivative of $dl(x,c)$ with regard to $x$. \\

\begin{center}

$dl'(x,c) = - \frac{2c^2 ln(x) - x^2 - 2c^2 ln(c) + c^2}{2(x+c)^2}$

\end{center}
We can rewrite this derivative to be in the form:

\begin{center}

$dl'(x,c) = \frac{2c^2 ln(x) - x^2 - 2c^2 ln(c) + c^2}{2(x+c)^2}$

\end{center}
Note that similarly when $x=c$ the second derivative  is equal to 0. With this understanding we take the second derivative.\\
 \begin{center}
$dl''(x,c) = \frac{c(x^2 + 2cxln(x) - 2cxln(c) -c^2)}{x(x+c)^3}$
\end{center}

Firstly, with $d''(x,c)$ we also know that when $x=c$, $d''(x,c) = 0$. \\

Secondly, when $x>c$ we know that $x$ and $c$ are both positive meaning that the denominator of $d''(x,c)$ is positive. We can rearrange the numerator to be in the form:

\begin{center}
    $c(x^2 - c^2 + 2cxln(\frac{x}{c})$

\end{center}

Since $x > c$ we know both $x^2 - c^2$ and $2cxln(\frac{x}{c})$ are postive and since by definition c is also positive, this means the numerator is also positive.\\

Thus the $d''(x,c)$ is always non-negative when $x \geq c$, and $d'(x,c)$ and $d(x,c)$ are 0 when $x=c$, we know both functions will be non-negative. \\

Thus, $dl(x,c)>0$.


Therefore, $DL(P,Q) \geq 0$

\subsection{4th Metric Property: Triangular Inequality}

\textbf{Theorem 2.} \emph{For any probability distributions P, Q, and R:}

\centering $\sqrt[3]{DL(P,Q)} + \sqrt[3]{DL(Q,R)} \geq \sqrt[3]{DL(P,R)}$

\raggedright
\emph{Proof.}

Suppose we have:
\begin{center}
$\sqrt[3]DL(P,R) = \sqrt[3]{\sum_{p_i,q_i}^{} x(1-ln(x)) - c(1-ln(c)) - \frac{x^2(1-2ln(x))-c^2(1-2ln (c))}{2(x+c)}} $
\end{center}
such that if $p_i>q_i$, we set $p_i = x_i$ and $q_i = c_i$ and if $q_i>p_i$, we set $q_i = x_i$ and $p_i = c_i$.
\\
Since all $dl(p_i,q_i)$ will be of this form, we simplify this to.
\begin{center}
$\sqrt[3]DL(P,R) = \sqrt[3]{m \cdot (x(1-ln(x)) - c(1-ln(c)) - \frac{x^2(1-2ln(x))-c^2(1-2ln (c))}{2(x+c)})} $

\end{center}
for some $m \in \mathbb{R}$.\\

The second derivative of this is:
\begin{center}
$-\frac{m^2\cdot\left(\left(12c^3x^2+4c^4x\right)\ln^2\left(x\right)+\left(-4c^2x^3-24c^3\ln\left(c\right)\,x^2+\left(4c^4-8c^4\ln\left(c\right)\right)x\right)\ln\left(x\right)+x^5-3cx^4\right)}{18x\cdot\left(x+c\right)^4\left(mx\cdot\left(1-\ln\left(x\right)\right)-\frac{m\cdot\left(x^2\cdot\left(1-2\ln\left(x\right)\right)-c^2\cdot\left(1-2\ln\left(c\right)\right)\right)}{2\left(x+c\right)}-c\cdot\left(1-\ln\left(c\right)\right)m\right)^\frac{5}{3}} + \frac{m^2\cdot\left(\left(4c^2\ln\left(c\right)-2c^2\right)x^3+\left(12c^3\ln^2\left(c\right)+6c^3\right)x^2+\left(4c^4\ln^2\left(c\right)-4c^4\ln\left(c\right)+c^4\right)x-3c^5\right)}{18x\cdot\left(x+c\right)^4\left(mx\cdot\left(1-\ln\left(x\right)\right)-\frac{m\cdot\left(x^2\cdot\left(1-2\ln\left(x\right)\right)-c^2\cdot\left(1-2\ln\left(c\right)\right)\right)}{2\left(x+c\right)}-c\cdot\left(1-\ln\left(c\right)\right)m\right)^\frac{5}{3}}$
\end{center}


The second derivative is always negative meaning it is concave. We also know from Theorem 1 that our function is always positive. Thus DLITE is subadditve, meaning it by definition satisfies the triangle inequality.

\subsection{Information-theoretic Property: Lemma for Ensemble as Weighted Sum}

\textbf{Lemma} \emph{Given the dl(p,q) function of probability change from p to q in Equation 11, for any positive value x:}

\centering $dl(xp,xq) = x \cdot dl(p,q)$

\raggedright
\emph{Proof.}

Let $p>q$, then $xp>xq$. \\

\begin{center}

$dl(xp,xq) = xp(1-ln(xp)) - xq(1-ln(xq)) - \frac{(xp)^2(1-2ln(xp)) - (xq)^2(1-2ln(xq)}{2(xq+xp)} $ \\

\bigskip
$= x(p(1-lnp) - q(1-lnq)) - xlnx(p-q) - x\frac{p^2(1-2lnp)-q^2(1-2lnq)}{2(p+q)} + \frac{2x^2p^2lnx - 2x^2q^2lnx}{2x(p+q)} $\\
\bigskip
$= x(p(1-lnp) - q(1-lnq))- x\frac{p^2(1-2lnp)-q^2(1-2lnq)}{2(p+q)}- xlnx(p-q)  + xlnx(p-q) $\\
\bigskip
$= x(p(1-lnp) - q(1-lnq))- x\frac{p^2(1-2lnp)-q^2(1-2lnq)}{2(p+q)} $\\
\bigskip
$= x \cdot ((p(1-lnp) - q(1-lnq))- \frac{p^2(1-2lnp)-q^2(1-2lnq)}{2(p+q)}) $\\
\bigskip
$= x \cdot dl(p,q)$
\end{center}

If $q>p$ the same result can be obtained.

\section{Conclusion}

DLITE functions as a metric space as it satisfies the four necessary proprieties, namely non-negativity, identity of indiscernibles, symmetry, and the triangle inequality. Satisfying these rules allows the ability to assume relationships between the amount of information in distributions based on their ``distance.'' There are more relationships in the theory of DLITE to be explored both mathematically and in practice.

\vspace{25em}
\bibliographystyle{abbrv}
\bibliography{dlite_report}

\begin{thebibliography}{10}

\bibitem{Aizawa:2003}
A.~Aizawa.
\newblock An information-theoretic perspective of tf--idf measures.
\newblock {\em Information Processing \& Management}, 39(1):45 -- 65, 2003.

\bibitem{Amati:2002}
G.~Amati and C.~J. Van~Rijsbergen.
\newblock Probabilistic models of information retrieval based on measuring the
  divergence from randomness.
\newblock {\em ACM Trans. Inf. Syst.}, 20(4):357–389, Oct. 2002.

\bibitem{Du2018}
Y.~Du, J.~Liu, W.~Ke, and X.~Gong.
\newblock Hierarchy construction and text classification based on the
  relaxation strategy and least information model.
\newblock {\em Expert Systems with Applications}, 100:157--164, 2018.

\bibitem{Ke2013JCDL}
W.~Ke.
\newblock Information-theoretic term weighting schemes for document clustering.
\newblock In {\em Proceedings of the 13th ACM/IEEE-CS Joint Conference on
  Digital Libraries}, JCDL '13, pages 143--152, New York, NY, USA, 2013. ACM.

\bibitem{Ke2015IJDL2}
W.~Ke.
\newblock Information-theoretic term weighting schemes for document clustering
  and classification.
\newblock {\em International Journal on Digital Libraries}, 16(2):145--159,
  2015.

\bibitem{Ke2017}
W.~Ke.
\newblock Text retrieval based on least information measurement.
\newblock In {\em Proceedings of the ACM SIGIR International Conference on
  Theory of Information Retrieval}, ICTIR '17, page 125–132, New York, NY,
  USA, 2017. Association for Computing Machinery.

\bibitem{Ke2020}
W.~Ke.
\newblock Dlite: The discounted least information theory of entropy, 2020.

\bibitem{Kullback:1959}
S.~Kullback.
\newblock {\em {Information Theory and Statistics}}.
\newblock Wiley, New York, 1959.

\bibitem{Robertson2009}
S.~Robertson and H.~Zaragoza.
\newblock The probabilistic relevance framework: {BM25} and beyond.
\newblock {\em Foundations and Trends{\textregistered} in Information
  Retrieva}, 3(4):333--389, 2009.

\bibitem{Shannon1948}
C.~E. Shannon.
\newblock A mathematical theory of communication.
\newblock {\em Bell System Technical Journal}, 27:379--423 and 623--656, July
  and October 1948.

\bibitem{Shaw:entropy}
D.~Shaw and C.~H. Davis.
\newblock Entropy and information: A multidisciplinary overview.
\newblock {\em Journal of the American Society for Information Science},
  34(1):67--74, 1983.

\end{thebibliography}

\end{document}